\documentclass [aps,prl,twocolumns]{revtex4}  
  
\begin{document}  
  
\begin{flushleft}  
LYCEN 2002-26  
\end{flushleft}  
  
\title{Relativistic dissipative hydrodynamics with spontaneous  
symmetry breaking.}  
\author{C. Pujol and D. Davesne}  
\address{IPN Lyon, 43 Bd du 11 Novembre 1918, F-69622 Villeurbanne Cedex}   
\date{\today}
\begin{abstract}  
In this paper we consider dissipative hydrodynamic equations for 
systems with continuous broken symmetries. We first present the case of 
superfluidity, in which the symmetry $U(1)$ is broken and then generalize to the
chiral symmetry $SU(2)_L \times SU(2)_R$. The corresponding new transport 
coefficients are introduced. 
\end{abstract}  
\maketitle  
  
\section{Introduction}  

Relativistic hydrodynamics has often been used as a starting point in the field 
of heavy ion collisions in order to reproduce single-particle spectra or 
multi-particle correlations (see \cite{Csernai} for example). However, the hot 
and dense hadronic phase created in such experiments is mainly constituted of 
pions and, until recently \cite{son}, no care has been taken on the fact that 
the breaking of the chiral symmetry can affect the theory of hydrodynamics 
itself \cite{moi,murongaWinter}. 
Among all the applications of such a modification of the theory, two main 
directions emerge~: first, it is highly wished to have a quantitative 
estimation of this effect when reproducing the experimental spectra \cite{nous} 
and, second, the aim of this paper, it is important to complete the whole 
theory, that is, to include dissipation.   
 
In hydrodynamic regime, relevant variables are those whose variations in space 
and time are slow and relaxation time becomes infinite in the long wavelength, 
small frequency limit. Such variables are of two kinds~: densities of conserved 
quantities and Goldstone modes associated to continuous broken symmetry 
\cite{forster}. The prototype of such a theory which includes both types of 
hydrodynamic variables is the well-established theory of superfluidity 
associated to the breaking of $U(1)$ and developed many years ago by Landau 
\cite{landau} in the non-relativistic case (see also \cite{carter,putt} in the 
relativistic domain). The example of the superfluidity will be used in the first
 part as a guideline in order to show explicitly how to include dissipation in 
hydrodynamic equations with a broken continuous symmetry. Then, in a second 
part, we will generalize to the case of $SU(2)_L \times SU(2) _R$~: starting
from the equations of ideal hydrodynamics, we will introduce explicitely new
transport coefficients associated to the chiral charge densities and Goldstone
modes.
  
\section{Superfluidity}   

Let us focus on our first example of hydrodynamics with a continuous broken
symmetry~: the superfluidity. As mentioned in the introduction, the spontaneous 
breaking of a continuous symmetry implies that the Goldstone modes are 
hydrodynamic variables, i.e. relax very slowly to equilibrium in the long 
wavelength, small frequency limit. Concerning the superfluid, one therefore has 
to introduce $\phi$, which is the phase of the condensate which breaks the 
$U(1)$ symmetry associated to the particle number, in our theory. More
precisely, the equations for ideal (that is without dissipation) relativistic 
superfluid can be written as \cite{son2}:   
\begin{eqnarray}  
& & \partial_\mu(n_0u^\mu-V^2\partial^\mu\phi)=0  \\
& & \hbox{\hskip 1 true cm} \partial_\mu T^{\mu\nu}=0  \\
& & \hbox{\hskip .8 true cm} u^\mu\partial_\mu\phi=-\mu_0 
\end{eqnarray}  
where $T^{\mu\nu}$ is the energy-momentum tensor, $V^2$ is the superfluid 
density in the non-relativistic limit and $\mu_0=\gamma \mu$ is the chemical 
potential ($\gamma$ is the Lorentz factor). We have to notice
at this stage that the third equation is equivalent to the one contained in the 
original two-fluid model of Landau \cite{landau} when making the correspondence~
:  $$\mu =\mu_{Landau}+\frac{v_s^2}{2}-\vec{v}_n.\vec{v}_s$$ 
where $\vec{v}_n$ is the spatial part of the four-velocity $u^\mu=\gamma 
(1, \vec v_n)$ and $\vec{v}_s = \frac{1}{m} \vec{\nabla}\phi$ is the superfluid 
velocity.

We can see that the presence of the new hydrodynamic variable $\phi$ manifests 
itself not only in the third equation but also in the first one (conservation of
 particle number) and in the second one since $T^{\mu\nu}$ is equal to~: 
\begin{equation}  
T^{\mu\nu}=(\epsilon + p)u^\mu u^\nu - p g^{\mu\nu} + V^2 \partial^\mu\phi
\partial^\nu\phi  
\end{equation}  
Since we are dealing with an ideal fluid the entropy conservation must be 
satisfied. Actually it is already contained in the second above equation when 
projected along the direction of $u^{\nu}$~:
$
u_{\nu}\partial_\mu T^{\mu\nu}=0 \hbox{  } \Rightarrow  \hbox{  }
\partial_\mu(s_0u^\mu)=0 .
$
 
Now, we can go further and add dissipative terms in the hydrodynamic equations.
The method, described for example in \cite{landau}, consists in introducing
fluxes such as~: 
\begin{eqnarray}  
& & \hbox{\hskip 1.5 true cm} \partial_\mu(n_0u^\mu-V^2\partial^\mu\phi
+\nu^\mu)=0  \\
& & \partial_\mu \left( (\epsilon + p)u^\mu u^\nu   
- p g^{\mu\nu} + V^2 \partial^\mu\phi \partial^\nu\phi +\tau^{\mu\nu} \right)=0
\\
& & \hbox{\hskip 2.2 true cm} u^\mu\partial_\mu\phi=-\mu_0-\phi'_0  
\end{eqnarray} 
Moreover, our choice for the hydrodynamic velocity $u^{\mu}$ imposes that 
$u_{\mu}  \tau^{\mu\nu}=0$ and $u_{\mu}\nu^\mu =0$ (the Eckart choice leads to 
other constraints, see \cite{weinberg} for example).

With these equations, we are now able to derive explicitely the new equation for
the entropy. Writing $u_{\nu}\partial_\mu T^{\mu\nu}=0$, we obtain~: 
\begin{equation}  
\partial_\mu(s_0u^\mu-\frac{\mu_0}{T_0}\nu^\mu)=-\nu^\mu\partial_\mu
\frac{\mu_0}{T_0}+\frac{\phi'_0}{T_0}\partial_\mu (V^2\partial^\mu\phi)
+\frac{\tau^{\mu\nu}}{T_0}\partial_\mu u_\nu  
\end{equation} 
This equation has the generic form~: $\partial_\mu S^{\mu} = \sigma$
where $S^{\mu}$ is the total four-flow entropy and $\sigma$ the entropy 
production. $\sigma$ is a bilinear form between fluxes and thermodynamic forces.
In the hydrodynamic regime, we are by definition near the global equilibrium so 
that we can express linearly the relation between these fluxes and these forces.
The coefficients of proportionality are called the transport coefficients. Their
 physical meaning is thus to characterize the magnitude of the response of the 
 system (flows) to a certain disturbance (thermodynamic forces).

Moreover, because of thermodynamics, $\sigma$ must be positive. This constraint 
when combined with the Onsager principle leads to~:
\begin{eqnarray}
& & \hbox{\hskip 5.5 true cm} \nu^\mu = \kappa(g^{\mu\nu}-u^\mu u^\nu)
\partial_\nu(\frac{\mu_0}{T_0}) \\
& & \tau^{\mu\nu} = (g^{\mu\nu}-u^\mu u^\nu)  
\left[\zeta_1\partial_\lambda(V^2\partial^\lambda\phi)+(\zeta_2-\frac{2}{3}\eta)
\partial_\lambda u^\lambda\right] + \eta \left[(g^{\mu\lambda}-u^\mu u^\lambda)
\partial_\lambda u^\nu+ (g^{\lambda\nu}-u^\lambda u^\nu )\partial_\lambda u^\mu)
\right] \\  
& & \hbox{\hskip 5.4 true cm} \phi'_0 = \zeta_1\partial_\mu u^\mu
+\zeta_3\partial_\mu(V^2\partial^\mu\phi)
\end{eqnarray}
where $\kappa$  is proportional to the thermal conductivity and $\eta,\zeta_1,
\zeta_2, \zeta_3$ are the shear and bulk viscosities (notations are those
defined in \cite{landau}). The positivity of $\sigma$ implies that $\eta,
\zeta_2, \zeta_3$ are positive and $\zeta_1^2 \le \zeta_2 \zeta_3$. The sign of 
$\zeta_1$ has to be determined by physical considerations only (here, it can be
checked by comparison with \cite{landau} that $\zeta_1$ is positive as well). 

\section{Chiral dynamics}  

\subsection{Ideal fluid}  
  
In this section, we are going to recall the main results of hydrodynamics with
chiral $SU(2)$ symmetry spontaneously broken \cite{son}. The hydrodynamic 
degrees of freedom in a chiral fluid are the densities of the conserved 
quantities, namely entropy density $s$, momentum density $T^{0i}$, baryonic 
number density $n$, left and right-handed charge densities written as 
$SU(2)$-matrices $\rho_L\equiv\rho_L^i\tau_i/2$ and $\rho_R\equiv\rho_R^i
\tau_i/2$ and finally the variables associated to the Golstone modes. For chiral
 symmetry, these modes are actually the pions. Then, following \cite{son}, we 
 can write the new hydrodynamic variables as a $SU(2)$-matrix $\Sigma \equiv 
 e^{i \vec \tau .\vec \pi / f_{\pi}}$ and, by analogy with the superfluid, 
 denote $\Sigma$ as "phases". The energy density $T^{00}$ is a function of all 
 these variables and of the first partial derivatives of $\Sigma$ (assuming 
 $\Sigma$ varies slowly). Following again \cite{son} we thus write the energy 
 density $T^{00}$ as~:
$$  
T^{00}=\epsilon_0 (s, n, T^{0i}) + \epsilon _1  
$$  
where $\epsilon_0$ is the "normal fluid" part and $\epsilon _1$ contains all 
the non trivial terms, at lowest order, compatible with chiral symmetry~:
\begin{eqnarray*}
\epsilon _1 & = & \frac{f_s^2}{4}(\delta_{ij}-\frac{1-v_{\pi}^2}{1-v_{\pi}^2v^2}
v^i v^j) \hbox{tr}\partial_i\Sigma \partial_j\Sigma^\dagger 
+ \frac{1}{\gamma ^2 f_t^2(1-v_{\pi}^2v^2)}\hbox{tr}(\rho_L-\Sigma\rho_R\Sigma^
\dagger)^2 \cr & & + \frac{1}{\gamma ^2 f_v^2}\hbox{tr}(\rho_L+\Sigma\rho_R
\Sigma^\dagger)^2 -i\frac{v_{\pi}^2}{\gamma ^2(1-v_{\pi}^2v^2)}v^k\hbox{tr}
(\rho_L  \Sigma\partial_k\Sigma^\dagger+\Sigma^\dagger\partial_k\Sigma\rho_R) 
\end{eqnarray*}
where $v_{\pi}\equiv f_s/f_t$ is the pion velocity and $f_s , f_t$ and $f_v$ are
functions to be determined by thermodynamics of the underlying fundamental
theory, namely QCD.

We have fourteen hydrodynamics variables. Therefore we must have fourteen
hydrodynamics equations. With the hamiltonian density above, it is possible to
show \cite{son} that the equations can be written in a covariant way only if we
make some combinations of the initial variables. Finally the result is~:
\begin{equation}  
\partial_\mu(n_0u^\mu)=0  
\end{equation}  
\begin{equation}  
\partial_\mu T^{\mu\nu}=0  
\label{tmunu}
\end{equation}  
\begin{equation}  
\partial_\mu(\alpha u^\mu)+\frac{1}{2}[A,\alpha]=0  
\label{alpha}
\end{equation}  
\begin{equation}  
i\partial_\mu((f_t^2-f_s^2)u^\mu A+f_s^2\Sigma\partial^\mu\Sigma^\dagger)+
[A,\alpha]=0  
\label{sigma}
\end{equation}  
with $T^{\mu\nu}=(\epsilon + p)u^\mu   
u^\nu - p g^{\mu\nu} + \frac{f_s^2}{4}\hbox{tr}(\partial^\mu\Sigma\partial^\nu 
\Sigma^\dagger + \partial^\nu\Sigma\partial^\mu\Sigma^\dagger)$ and 
$A=u^\mu\Sigma\partial_\mu\Sigma^\dagger$.

The nine equations which are specific to chiral dynamics ($\partial_\mu J^\mu_
{L,R}=0$ and  the first order equation for $\Sigma$) are actually contained in 
the first order equation (\ref{alpha}) for $\alpha\equiv (\rho_L+\Sigma\rho_R
\Sigma^\dagger)/\gamma$ and in the second order equation (\ref{sigma}) for 
$\Sigma$ constructed from the combination $(\rho_L-\Sigma\rho_R\Sigma^\dagger)$.

Once again, starting from equation (\ref{tmunu}) and using all the other 
(hydrodynamic and thermodynamic) equations, it is possible to deduce the 
entropy conservation~: $\partial_\mu(s_0u^\mu)=0$ as it should be for a perfect
fluid.
At this stage, we can already make some comparisons with the superfluid. First 
of all if $\Sigma$ was associated with the breaking of $U(1)$ type symmetry,
namely $\Sigma=e^{i\phi}$ , then $i\Sigma\partial_\mu\Sigma^\dagger$ would be
equal to $\partial_\mu\phi$. This means that $i\Sigma\vec{\nabla}\Sigma^\dagger$
plays the role of superfluid velocity. Then, by direct comparison between
the energy-momentum tensors (for $U(1)$ and $SU(2)$), we see that 
$\frac{f_s^2}{2}$ can be interpreted as "superfluid density".
  
\subsection{Dissipation}  
  
We are now in position to treat the dissipation case. As for the superfluid, we 
add some dissipative flux densities to conservation equations~:  $\nu^\mu$ for 
baryonic number, $j^\mu_{L,R} \equiv j^\mu_{L,R;i} \tau _i /2$ for chiral
charges and $\tau^{\mu\nu}$ for energy-momentum tensor; and we add also a 
dissipative term for the other
non-conserved variables~: $\frac{i}{2}\Sigma_0' \equiv \frac{i}{2}\Sigma_{0,i}'
\tau _i /2$.  

Then, performing the same combinations as in the previous section, we can 
finally write after some calculations~:   
\begin{equation}  
\partial_\mu(n_0u^\mu+\nu^\mu)=0  
\end{equation}  
\begin{equation}  
\partial_\mu \left((\epsilon + p)u^\mu  u^\nu - p g^{\mu\nu} + 
\frac{f_s^2}{4}\hbox{tr}(\partial^\mu\Sigma\partial^\nu\Sigma^  
\dagger + \partial^\nu\Sigma\partial^\mu\Sigma^\dagger)+\tau^{\mu\nu} \right)=0 
\end{equation}  
\begin{equation}  
\partial_\mu(\alpha u^\mu)+\frac{1}{2}[A^{(0)},\alpha]=
-\partial_\mu j^\mu_L-\Sigma \partial_\mu j^\mu_R\Sigma^\dagger  
\end{equation}  
\begin{equation}  
i\partial_\mu\left((f_t^2-f_s^2)u^\mu A^{(0)} +
f_s^2\Sigma\partial^\mu\Sigma^\dagger\right) +[A^{(0)},\alpha]=
2\partial_\mu j^\mu_L-2\Sigma \partial_\mu j^\mu_R\Sigma^\dagger  
\end{equation}  
with $A=A^{(0)}+\frac{i}{2}\Sigma_0'$ which is the dissipative first order
equation
for $\Sigma$. $A^{(0)}$ represents the non-dissipative part of $A$. 
With these new equations, it is a simple task to express the equation for the 
entropy (note that, since the expressions for $\epsilon$ and
$p$ must be given by the same formula as in the non-dissipative case, $\epsilon$
and $p$ depend actually only on $A^{(0)}$ and not on $A$)~:
\begin{eqnarray}
\partial_\mu\left(s_0u^\mu-\frac{\mu_0}{T_0}\nu^\mu - \hbox{tr}
\left(\frac{\mu _{L0}}{T_0}j^\mu_L 
+ \frac{\mu _{R0}}{T_0} j^\mu_R \right)\right)&=& -\nu^  
\mu\partial_\mu(\frac{\mu_0}{T_0})
+\frac{\tau^{\mu\nu}}{T_0}\partial_\mu u_\nu \nonumber\\
&-& \hbox{tr} \left(
j^\mu_L\partial_\mu\left(\frac{\mu _{L0} }{T_0}\right)+
j^\mu_R\partial_\mu\left(\frac{\mu_{R0}}{T_0}\right)+
\frac{\Sigma_0'}{2T_0}\partial_\mu\left(\frac{f_s^2}{2}i\Sigma\partial^\mu\Sigma^
\dagger\right)\right)
\end{eqnarray}
where $\mu _{L0} = {\displaystyle \frac{2\alpha}{f_v ^2}}-iA^{(0)}$ and  
$\mu _{R0} = \Sigma^\dagger ({\displaystyle \frac{2\alpha}{f_v
^2}}+iA^{(0)})\Sigma$.

The introduction of the above shorthand notations $\mu _{R0}$ and $\mu _{L0}$
is motivated by the fact that they can be identified as chemical potentials 
but for the left
and right chiral densities (they are thermodynamic conjugate variables of 
$\rho_L$ and $\rho_R$). Nevertheless, due to the fact that we are considering 
the $SU(2)$ case and no more $U(1)$, new features appear, that is we have now some
couplings with the derivatives of these chemical potentials $\mu _{R0}$ and
$\mu _{L0}$.

On the right-hand side of the equation, the entropy
production is, as usual, a bilinear form between "thermodynamic forces" and
dissipative fluxes. Again, we can make linear combinations with those
thermodynamic forces and introduce transport coefficients. Our choice  
for the hydrodynamic velocity imposes also that $u_{\mu}j^\mu_{L}=0$ and 
$u_{\mu}j^\mu_{R}=0$ and we  have to make the entropy production positive. These
prescriptions allow us to eliminate some coefficients and to obtain some
constraints on the remaining others. We get~:  
\begin{equation}  
\Sigma_0'=\left(\zeta_{4,i}\partial_\mu u^\mu+[\zeta_3]_{i,j}\partial_\mu
\left(\frac{f_s^2}{2}i(\Sigma\partial^\mu\Sigma^
\dagger)_j\right)\right)\tau _i
\end{equation}  
\begin{equation}  
\nu^\mu=(g^{\mu\nu}-u^\mu  
u^\nu)\left(\kappa\partial_\nu\left (\frac{\mu_0}{T_0}\right)+\kappa_{L,i}.
\partial_\nu\left (\frac{\mu _{L0,i}}{T_0}\right)+
\kappa_{R,i}.\partial_\nu\left(\frac{\mu _{R0,i}}
{T_0}\right)\right)  
\end{equation}  
\begin{equation}  
j^\mu_{L}=(g^{\mu\nu}-u^\mu  
u^\nu)\left(\kappa_{L,i}\partial_\nu\left (\frac{\mu_0}{T_0}\right)
+[\kappa_{LL}]_{i,j}
\partial_\nu\left(\frac{\mu_{L0,j}}{T_0}\right)+[\kappa_{LR}]_{i,j}
\partial_\nu\left(\frac{\mu_{R0,j}}{T_0}\right)\right)\tau _i
\end{equation}  
\begin{equation}  
j^\mu_R=(g^{\mu\nu}-u^\mu  
u^\nu)\left(\kappa_{R,i}\partial_\nu\left (\frac{\mu_0}{T_0}\right)
+[\kappa_{LR}]_{j,i}
\partial_\nu\left(\frac{\mu_{L0,j}}{T_0}\right)+[\kappa_{RR}]_{i,j}
\partial_\nu\left(\frac{\mu_{R0,j}}
{T_0}\right)\right)
\tau_i  
\end{equation}  
\begin{equation}  
\tau^{\mu\nu}=(g^{\mu\nu}-u^\mu
u^\nu)\left((\zeta_2-\frac{2}{3}\eta)\partial_\lambda u^\lambda +
\zeta_{1,j} \partial_\lambda
\left(\frac{f_s^2}{2}i(\Sigma\partial^\lambda\Sigma^
\dagger)_j\right)\right)  
+\eta\left((g^{\mu\lambda}-u^\mu u^\lambda)\partial_\lambda u^\nu+  
(g^{\lambda\nu}-u^\lambda u^\nu)\partial_\lambda u^\mu\right)  
\end{equation}  
where $[Q]$ means that $Q$ is a 3$\times$3 matrix. Due to the Onsager reciprocity
principle, all the matrices except $[\kappa_{LR}]$ are symmetric, $4\zeta_{1,i}=
\zeta_{4,i}$ and there are actually 39 independant
coefficients. If we represent the quadratic form of the entropy production by a
12 $\times$ 12 matrix, we can easily show that all coefficients appearing in the
diagonal should be positive and that there exists inequalities between the 39
coefficients~: as for the superfluid, all the principal minors have to be
positive. 
We see that dissipative equations for $SU(2)_L \times SU(2)_R$ symmetry implies 
new transport coefficients and couplings between baryonic, left-handed and 
right-handed currents. We also see that spontaneous breaking of the chiral 
symmetry implies the existence of matricial transport coefficient for, by
instance, the bulk 
viscosities $\zeta_1$ and $\zeta_3$~: the thermodynamic force $\partial_\lambda
\left(\frac{f_s^2}{2}i(\Sigma\partial^\lambda\Sigma^\dagger)_j\right)$ which is the
equivalent of $\partial_\lambda(V^2\partial^\lambda\phi)$ has now three
components. These components are nevertheless all multiplied by the same
factor $\frac{f_s^2}{2}$ which is related to the amplitude of the order
parameter of the chiral symmetry. Of course, near the transition phase this
amplitude is not frozen as in our case and can fluctuate. The dynamics of these
fluctuations is such that the amplitude of the order parameter has to be
incorporated explicitely in the approach since then it becomes an hydrodynamic
variable \cite{Ginzburg}. 

Finally let us remark that, since it is known that dissipation can
affect the observables (see \cite{muronga} for the modification of the
temperature profile used to describe heavy ion collisions), it will be of fundamental
importance to determine quantitatively the influence of these new couplings.

\section{Conclusions and outlook}  

We have treated in this paper the dissipation in the hydrodynamic regime for
relativistic systems with spontaneously broken symmetry $U(1)$ and $SU(2)$. For 
the superfluid, we recovered the non-relativistic limit of \cite{landau}. For 
the $SU(2)$ case, we introduced new transport coefficients associated to the 
right and left charges and to the Goldstone modes. Since we know that transport
coefficients can affect quantitatively observables, it would be interesting to
express them with Kubo-type relations and then compute them explicitely from
this microscopic approach. This work is under study \cite{renous}. An other extension of this
paper is to determine the new relaxation times with the effect of the symmetry
breaking in order to know if this implies some noticeable modifications to
the conclusions drawn in \cite{moi} concerning the typical equilibration time of heavy ions
collisions. It is well known that the way we introduced the
dissipative effects in that paper leads formally to instabilities and can not
allow to reach the underlying physics of the relaxation times. In order to
solve that problem one has to introduce them explicitely. One
convenient way to do this is to use the 14-moments of Grad as done in \cite{moi}.

\vglue 1 true cm

Acknowledgments~: We thank G. Chanfray and D.T. Son for useful discussions.

\end{document}